\documentclass[10pt,conference]{IEEEtran}

\usepackage[utf8]{inputenc}
\usepackage{booktabs} % used to configure tables
\usepackage{multirow} % used to configure tables
\usepackage{multicol} % used to configure tables
\usepackage{pifont}
\usepackage[caption = false]{subfig}
\usepackage{ifpdf}
\usepackage{amsmath}
\usepackage{algorithmic}
\usepackage{array}
\usepackage{color}
\usepackage{listings}
\usepackage{graphicx}
\usepackage{xcolor}
\usepackage[utf8]{inputenc}
\usepackage[T1]{fontenc}
\usepackage[hyphens]{url}
\usepackage{stfloats}
\usepackage{url}
\usepackage{array}
\usepackage{booktabs}
\usepackage{textcomp}
\usepackage{multirow}
\usepackage{mathtools}
\usepackage{mathabx}
\usepackage{booktabs}
\usepackage{tikz}
\usepackage{pgf,tikz}
\usetikzlibrary{shapes.arrows,chains,calc}
\usepackage{color}
\usepackage{latexsym}
\usepackage{algorithm}
\usepackage{algorithmic}

\newcommand{\tikzcircle}[2][red,fill=red]{\tikz[baseline=-0.5ex]\draw[#1,radius=#2] (0,0) circle ;}%

\IEEEoverridecommandlockouts
\IEEEpubid{\makebox[\columnwidth]{978-1-5386-2524-8/17/\$31.00~\copyright2017 IEEE \hfill} \hspace{\columnsep}\makebox[\columnwidth]{ }}

\begin{document}

	\title{Personal Virtual Traffic Light Systems}
	\author{
		\IEEEauthorblockN{
            Vanessa Martins\IEEEauthorrefmark{1},
			Jo\~{a}o Rufino\IEEEauthorrefmark{1},
			Bruno Fernandes\IEEEauthorrefmark{1},
            Luís Silva\IEEEauthorrefmark{1},
            Jo\~{a}o Almeida\IEEEauthorrefmark{1}$^{,}$\IEEEauthorrefmark{2},
			Joaquim Ferreira\IEEEauthorrefmark{1}$^{,}$\IEEEauthorrefmark{3},
            Jos\'{e} Fonseca\IEEEauthorrefmark{1}$^{,}$\IEEEauthorrefmark{2},
		}
		\IEEEauthorblockA{\IEEEauthorrefmark{1} Instituto de Telecomunica\c{c}\~{o}es, Campus Universit\'{a}rio de Santiago, 3810-193 Aveiro, Portugal}
        		\IEEEauthorblockA{\IEEEauthorrefmark{2} DETI - Universidade de Aveiro, Campus Universit\'{a}rio de Santiago, 3810-193 Aveiro, Portugal}
			\IEEEauthorblockA{\IEEEauthorrefmark{3} ESTGA - Universidade de Aveiro, 3754-909 \'{A}gueda, Portugal}	
		\{vamartins,joao.rufino,brunofernandes lems,jmpa,jjcf,jaf\}@ua.pt
	}

	\maketitle
	
	\begin{abstract}
Traffic control management at intersections, a challenging and complex field of study, aims to attain a balance between safety and efficient traffic control. Nowadays, traffic control at intersections is typically done by traffic light systems which are not optimal and exhibit several drawbacks, e.g. poor efficiency and real-time adaptability. With the advent of Intelligent Transportation Systems (ITS), vehicles are being equipped with state-of-the-art technology, enabling cooperative decision-making which will certainly overwhelm the available traffic control systems. This solution strongly penalizes users without such capabilities, namely pedestrians, cyclists and other legacy vehicles. Therefore, in this work, a prototype based on an alternative technology to the standard vehicular communications, BLE, is presented. The proposed framework aims to integrate legacy and modern vehicular communication systems into a cohesive management system. In this framework, the movements of users at intersections are managed by a centralized controller which, through the use of networked retransmitters deployed at intersections, broadcasts alerts and virtual light signalization orders. Users receive the aforementioned information on their own smart devices, discarding the need for dedicated light signalization infrastructures. Field tests, carried-out with a real-world implementation, validate the correct operation of the proposed framework.
	\end{abstract}
	

\IEEEpeerreviewmaketitle
	
	\section{Introduction}
Transportation has always been a fundamental component of the economic and social interactions of the human society; without it, the movement of people, animals and goods from one geographical location to another would be highly constrained. From domesticated animals and wheel carts, to wheeled motor vehicles and airplanes, transportation means and infrastructures are in perpetual evolution, with significant impacts on the economy and environment.

In the last decades the number and density of vehicles, especially on road traffic, has increased significantly. In 2010, the number of vehicles operating worldwide surpassed the one thousand million units; this number roughly equates to a 1:6.75 ratio of vehicles to people in a world population of 6.9 billion \cite{wardsautovehicles}. According to \cite{statistavehicles}, the production and sale of automobiles have reached the 70 million units in 2014, a notable increase from the 50 million units sold between 2000 and 2013. A sustained raise of these numbers is foreseen for the next years due to the development of highly populated countries such as India.

The massive increase in both road traffic and population lead to an increase of accidents and congestion, along with negative impacts on the economy, environment and in the quality of people’s lives. According to the World Health Organization (WHO), road traffic injuries are expected to be the leading cause of death for people aged 15-29; projections show that road traffic deaths will become the seventh main cause of death worldwide by 2030 \cite{whomortality}. Despite being considered the safest, a report from the European Commission reveals that in 2015 nearly thirty thousand lives were lost in European roads \cite{ecroadsafety}. Although deaths in European roads have steadily decreased over the years, progress has slowed down, with the change in fatality figures close to zero from 2013 to 2015. When comparing the number of fatalities in high-ways, rural, and urban roads, statistics show that the majority of occurrences are in rural environments (55\%), followed by urban areas (37\%). Regarding the parties involved in accidents, 46\% of the accounted are automobile drivers, followed up by pedestrians (21\%). The latter are the most vulnerable road users; current efforts and measures have been struggling to reduce the number of victims in this group \cite{ecroadsafety}.

In order to increase the efficiency of road traffic and security for its users, new systems and applications are being researched. Intelligent Transportation Systems (ITS), sprouted by technological advancements in communication networks and computer science, comprise a set of these new applications and is currently under heavy debate and work by governmental organizations and scientific communities. In these systems, it is envisioned that vehicles and infrastructures deployed along roads may use wireless communications to exchange data between themselves and other users, such as pedestrians carrying smart devices (e.g. smartphones and watches). Several types of data can be exchanged, ranging from hazard alerts to geo-related data and infotainment.

Traffic control management at intersections, a challenging and complex field of study, is one of the applications targeted by ITS. Intersections require a balance between safety and efficient traffic control: in order to reduce congestion, the number of vehicles and users passing through an intersection should be maximized without compromising the security of the involved parties. Nowadays, traffic at critical intersections is typically managed through traffic light signalization. Albeit being able to manage traffic with significant success in terms of security to its users, this solution is not optimal and exhibits several drawbacks: (i) light signalization is expensive and thus, only a small percentage of all intersections are covered, (ii) adaptability to real-time conditions, such as the number of vehicles in a given lane, is poor and leads to an inefficient management of the volume of users crossing by: for example, light signalization may prioritize the crossing of an "empty" lane while other lanes are congested with users waiting for their turn to cross.

Due to the aforementioned issues, several proposals have been presented to improve traffic control management at intersections. A significant number of these solutions are based on future vehicular communication systems and do not address the integration of legacy systems, systems which, are expected to coexist for a significant amount of time\cite{jchang_marketpen} and thus, should not be ignored.

In this work, a solution based on an alternative technology to vehicular communications, Bluetooth Low Energy, is presented. This solution discards the need of an existing light signalization infrastructure and aims to control all intersection users by displaying semaphoric information in legacy in-vehicle displays and/or smart devices. Moreover, the proposed framework aims to integrate legacy and modern vehicular communication systems into a cohesive management system. The proposed platform is based on low cost, commercial-of-the-self hardware so that adoption and deployment in urban areas is feasible. The feasibility of the presented solution is validated with field tests carried-out with a real-world implementation.

The rest of the paper is organized as follows: Section \ref{sec:rework} depicts the state of the art and \ref{sec:tls} exposes the main characteristics of Traffic Light Systems, sections \ref{sec:arch} and \ref{sec:implementation} present the proposed framework and implementation, named Personal Virtual Traffic Light Systems. In Section \ref{sec:experimental_eval} the experimental and obtained results are shown and discussed. Finally, section \ref{sec:conclusions} presents the conclusion and future work.

\section{Related work}
\label{sec:rework}

As previously discussed, there have been efforts in the literature to improve traffic control management at intersections. The most relevant for this work are going to be briefly discussed in this section.

Michel et al.\cite{Ferreira:2010:STC:1860058.1860077} propose a virtual traffic light management framework that dynamically optimizes traffic flows in road intersections without requiring any roadside infrastructure. Vehicles use standard ITS Dedicated Short-Range Communication (DSRC) to elect a leader vehicle at the intersection that acts as a coordinator and broadcasts virtual traffic light messages to vehicles in range. These instructions are shown to drivers through in-vehicle displays. Simulation results show an increase in traffic efficiency with the proposed system. Hugo et al.\cite{Conceicao_2013} propose to equip DSRC and Virtual Traffic Lights (VTL) equipped vehicles with an exterior representation, in the form of a visible light, of their VTL rules, i.e. cross or stop. The purpose of this light is to guide the behavior of legacy car drivers, that should adjust accordingly. Simulations using traffic simulators and the network simulator NS3 show an increase in efficiency, with less trip duration times for scenarios with equipped vehicles. Nakamurakate et al.\cite{Nakamurakare_2013} developed a VTL framework for Android-based smartphones that conveys traffic light information to its users. The framework is based on a self-organized traffic control paradigm that detects and resolves conflicts at intersections. Vehicles use WiFi Direct ad-hoc communications to participate in a leader election process upon intersections. The elected leader then acts as a temporary traffic light infrastructure and broadcasts traffic light rules to nearby vehicles. The received information is then showed to the user through the smartphone's VTL application interface. A. Bazzi et al.\cite{Bazzi_2014} designed a virtual traffic light (VTL) in which a distributed algorithm exploits Vehicle-to-Vehicle (V2V) communications, based in IEEE 802.11p wireless communications, to broadcast messages that convey priorities to the vehicles approaching intersections. The algorithm has been tested in both laboratory and field trial environments. I. Iglesias et al.\cite{Iglesias_2008} propose a driving assisting system, based on Infrastructure-to-Vehicle (I2V) communications, that collects information regarding the speed and location of vehicles in the vicinity of an intersection. Infrastructure retrieves information regarding the current status of traffic lighting signalization using Wi-Fi communications. The system notifies drivers of the current traffic light state at the intersection through a in-vehicle display. It also combines all the gathered information to predict traffic light states at the instant in which drivers would reach the intersection. 

Works exploiting technologies besides standard ITS communication technologies also exist. R. Gheorghiu et al.\cite{Gheorghiu_2016} evaluated the use of ZigBee technologies for Vehicular-to-Infrastructure (V2I) communications. The system would be installed in emergency vehicles that, when in the vicinity of an intersection, would communicate with the traffic light system so the latter could facilitate the crossing of the emergency vehicle. The platform was validated using a laboratory testbed. V. Kodire \cite{Kodire_2016} proposed a similar system with ZigBee and V2I communications for crossing preemption of emergency vehicles at intersections. However, this system also relies on GPS information to increase efficiency.

The existing solutions differ from the work presented in this paper since none exploits Bluetooh Low Energy (BLE)~\cite{bluetooth_v4_0} as an alternative technology. Moreover, our work focus on a centralized approach that combines both legacy and ITS systems; the aforementioned works either employ distributed approaches \cite{Ferreira:2010:STC:1860058.1860077,Conceicao_2013,Nakamurakare_2013,Bazzi_2014}, discard the support for legacy vehicles \cite{Iglesias_2008}, or focus on applications oriented towards emergency vehicles \cite{Gheorghiu_2016,Kodire_2016}.

\section{Traffic Light Systems}
\label{sec:tls}

In order to propose a new architecture, we should first study the main characteristics of the conventional traffic control systems (TCS). Typically, there are four different roles in any TCS: users, data collectors, indication providers and controllers. A user is any actor receiving a traffic signal service such as, for example, pedestrians, cyclists and vehicles. Strategically positioned sensors have the role of detecting users and providing real-time information on the surrounding environment, e.g. pressure sensors and buttons for the detection of vehicles and pedestrians. In order to properly alert users on their next movements, alerting systems can either be physically deployed near intersections or presented before hand using some sort of smart-device. Finally, the controller is a decision-making agent responsible for the coordination of signaling devices. Controllers can be divided into two main groups: aware and unaware of the surrounding environment status. 

%%%%%%%%%%%%%%%%%%%%% Descrição dos unaware "pobre" %%%%%%%%
The first group offers greater flexibility and adaptability since they use the information provided by data collectors and other systems to conduct the decision process. In comparison, the latter solution is mostly used in zones with a significant amount of intersections, obligating a timely coordination of their status.
%%%%%%%%%%%%%%%%%%%%%%%%%%%%%%%%%%%%%%%%%%%%%%%%%%%%%%%%%

Despite being a small set of signaling devices, traffic lights are widely used due to their unobtrusive and intuitive operation. The set of colors currently being displayed by a traffic light is called a phase; movements are referred as the users' possible actions. Conventionally, these systems are based on 4 different colors: users are permitted to move if the light being displayed is green, and should halt on the presence of a red light. The yellow is presented when a status change is imminent,  e.g. from green to red, and the orange color is used for temporary traffic control. Moreover, when the user has the right-of-way he is conducting a protected movement. However, if the light is green and he has to yield to other users the movement is called permitted. Using these concepts, an intersection can be modeled by consistently numbering all movements and phases available\cite{urbanik2015signal}. One example is shown in Figure \ref{fig:phases} where a four-way intersection is thoroughly depicted. The model is based on the following rules:

\begin{figure}[htb!]
 	\centering
 	\includegraphics[width=\columnwidth]{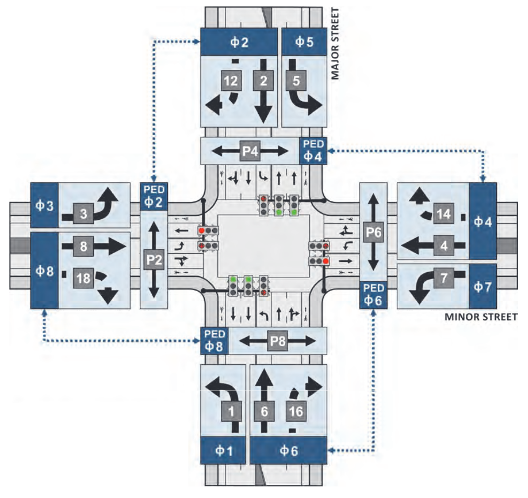}
 	\caption{Phases scheme (adapted from \cite{urbanik2015signal})}
    \label{fig:phases}
 \end{figure}
 
\begin{itemize}
\item Numbers, or in case of pedestrians a "P" followed by a number (Px), describe possible user movements at the intersection. These are represented in Figure~\ref{fig:phases} as gray squares;
\item In a four-way intersection, vehicles can perform twelve one-way movements while pedestrians can perform four two-way movements;
\item A traffic signal phase, $\Phi x$, represents a time period managed by a signal controller authorizing one or more movements at each intersection segment. Signal phases are depicted in Figure~\ref{fig:phases} as blue boxes;
%%%%%%%% O QUE É UM MOVIMENTO MAJOR THROUGH? %%%%%%%%%
\item Even numbers are used to reference go through and right-turn movements. Even numbers higher than 10 reference right permitted movements. Numbers 2 and 6 are used for major street trough movements;
\item Odd numbers reference left-turns. The numbers 1 and 5 are used for major street left turns; numbers 3 and 7 indicate minor street left-turn movements.
\item Pedestrian movements occur concurrently with even numbered vehicular phases, as depicted in dashed lines.
\end{itemize}
%%%%%%%%%%%%%%%%%%%%%%%%%%%%%%%%%%%%%%%%%%%%%%%%%%%%%%%%

\section{Proposed Architecture}
\label{sec:arch}
As discussed in section \ref{sec:tls}, vehicles are one of the multiple actors in the complex road safety system. Despite public and private efforts, there is still a long time to go until autonomous and cooperative vehicles make the use of nowadays physical traffic lights obsolete. In these futuristic systems, there is an extreme disregard for the role of pedestrians, cyclists and other non-cooperative systems; these are seen as a liability to themselves and to others. In our proposal we follow a non-elitist approach, empowering these users with an application that can be deployed in their most used smart device. This procedure makes our proposal unique since these devices are seen as a significant source of distraction for both drivers and pedestrians and consequently, a factor that may increase pedestrian fatalities \cite{smartphoneBBC,smartphoneGHSA}. However, we expect to highly reduce these prejudicial outcomes by fully exploiting its capabilities in order to provide precise, useful information in a clear and attention-grabbing manner. %%%%DEVIAMOS ARRANJAR ALGUM EXEMPLO....SINALIZAR PERIGO COM TOQUES/VIBRAÇAO CONSTANTE? DUNNO....

\begin{figure*}[!h]
 	\centering
 	\includegraphics[width=0.8\linewidth]{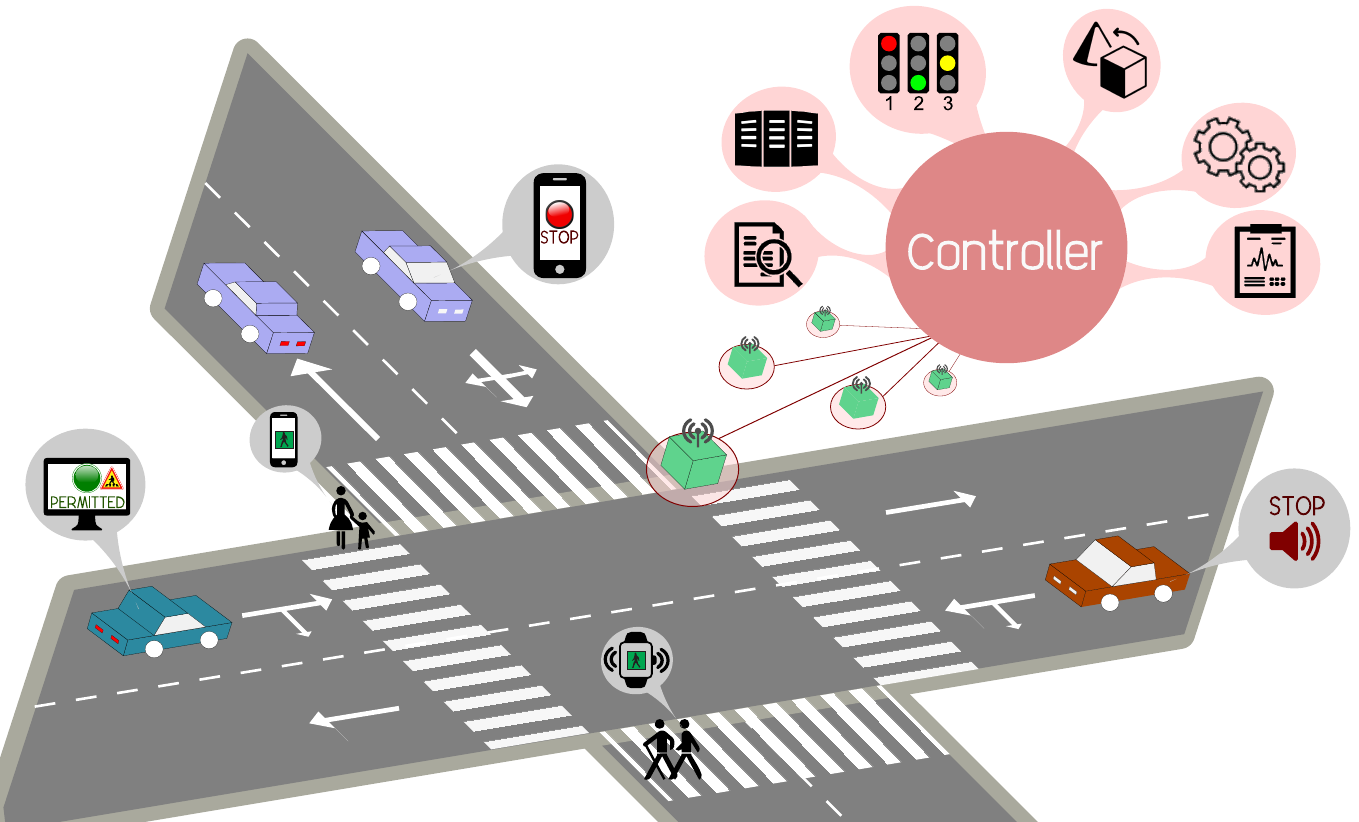}
 	\caption{Overview of the proposed system architecture}
    \label{fig:arch}
 \end{figure*}

An overview of the proposed architecture is depicted in Figure \ref{fig:arch}. A centralized controller holds information about the status of the traffic light system. Stored information includes the past, current and future states of the lightning system, as well as the time left until the next phase change. The controller may also be connected to other sources of information, such as road sensors, radars, cameras, etc. These inputs could be useful in order to enable a more precise and dynamic control of the intelligent intersection, crosswalk or roundabout. However, in this work the scenario will be restricted to a static scheme with pre-established phase sequences.

Typically, there is only one logical central controller, comprised of one or a limited number of synchronized nodes, responsible for the simultaneous coordination of all traffic lights in a city. Since the number of coordinating nodes is usually small, controllers fall short on delivering the information to every actor on the network. In the proposed architecture we use relay agents, called retransmitters, to extend the communication range and improve the delivery rate of information. These retransmitters work as mediators between actors and the controller, receiving the information, interpreting it and either delivering it to the controller or broadcasting it to the medium. Retransmitters are deployed in strategic positions, offering reliable connection points near sensitive spots, e.g. at low visibility and dense traffic areas, while also ensuring that users receive the information with sufficient reaction time to complete their action.

Figure \ref{fig:arch} presents the most common scenario where all road users are prompted before-hand with their permitted movements for the next intersection. A vehicle should be prompted to stop if pedestrians have the right-of-way. Moreover, an unaware pedestrian approaching a crossroad can be alerted through the vibration of its phone or by a sound signal emitted through his wireless headphones.

\subsection{Users}

In our proposal, an user is any sort of actor receiving information on the approaching intersection. Ideally, in order to improve the overall decision-making process, users are obliged to provide their own location. This is a deciding factor since it excludes the need for a vehicle and pedestrian detection. Clearly, there is a need for a precise localization mechanism, which in the future will certainly be provided with the 5G promising features. For the present, a combination of GPS coordinates and distance to nearby devices, e.g. retransmitters and other actors, should be used. 

% a localizacao é extremamente importante, permite que diferentes utilizadores
\subsection{Environment Perception}
Regarding environment perception, controllers are a meeting point for heterogeneous data sources, ranging from vehicular communications to multiple wireless sensor networks. Therefore, controllers have a global view of the environment and are able to devise intelligent management strategies. Even though there is redundancy at the controller level, mechanisms should be implemented in the case of communication loss and/or device failure. Ideally, retransmitters could also exploit their local environment perception since they work as mediators and receive information to/from the controller. In this approach, a leader can be elected to work as a temporary controller, receiving the information of the affected retransmitters and deploying
traffic control strategies within the coverage area. However, in the presented scenario we don't address controller failures and opt to simply notify actors of the connection failure.

\subsection{Signaling}
Smart devices have become workaday tools, commonly used in the almost everyday situation. In fact, we believe these devices can enhance or even replace the cumbersome, currently available, physical light systems. Their capabilities are unique since they provide direct communication with the users in the most comfortable and unobtrusive way. Thus, they are used to improve users awareness of the surrounding environment. Different alerting mechanisms are considered, namely, the use of their display as information provider, vibration as remainders of the distance to an intersection or emitting sound as an alerting signal.

Despite receiving the same information from the retransmitter, drivers and pedestrians shouldn't be alerted in the same form. In fact, there are different requirements for the alerting process. The interaction with drivers must be performed in such a way that reduces the impact on the driving process while clearly and intuitively present the status of each lane. The number of lanes is also a determining factor since in worst-case scenarios a pedestrian is presented with the status of two crossroads, while a vehicle can perform up to three different movements. In case of communication loss the smart-device should display a yellow sign recommending the user to proceed with caution at the next intersection.
%multiplos sensors disponivis

\subsection{Traffic control}

The controller is responsible for the proper functioning of the system. Together with the information gathering system, controllers combine historical data and exploit the rich information gathered in real-time from multiple sources, e.g. pressure sensors and traffic cameras, to efficiently adapt the traffic control strategies to the current traffic conditions. These units can also be connected to the available traffic light systems in order to orchestrate the lights in accordance to the alerts being emitted by smart devices. In our proposal, the traffic light phases are mapped into a matrix, as shown in Table \ref{tab:semaphore_states}, and the resulting matrix passed to retransmitters for dissemination. Upon transmission, relay nodes add a retransmitter tag to allow smart-devices to identify the source of the information.

\subsection{Communication}
There are a few requirements that come with the usage of smart devices. Firstly, the power consumption has to be low in order to reduce the battery drainage and secondly, it has to be compatible with most mobile phones, tablets, watches and other smart solutions. This confines us to a reduced array of communication technologies, being the most suited Bluetooth and Wi-Fi. BLE~\cite{bluetooth_v4_0} is perfectly suited for low-powered applications with a need to exchange simple bits of data between devices without pairing or any manual connection steps and thus, the opted choice for our platform.

% Bug IET template cannot use cline{...}, etc.
\begin{table*}[!htb]
\centering
\caption{Traffic Light temporal states}
\label{tab:semaphore_states}
{\begin{tabular}{>{\centering\arraybackslash}m{0.6cm}>{\centering\arraybackslash}m{0.6cm}>{\centering\arraybackslash}m{0.6cm}>{\centering\arraybackslash}m{0.6cm}>{\centering\arraybackslash}m{0.6cm}>{\centering\arraybackslash}m{0.6cm}>{\centering\arraybackslash}m{0.6cm}>{\centering\arraybackslash}m{0.6cm}>{\centering\arraybackslash}m{0.6cm}>{\centering\arraybackslash}m{0.6cm}>{\centering\arraybackslash}m{0.6cm}>{\centering\arraybackslash}m{0.6cm}>{\centering\arraybackslash}m{0.6cm}>{\centering\arraybackslash}m{0.6cm}>{\centering\arraybackslash}m{0.6cm}>{\centering\arraybackslash}m{0.6cm}>{\centering\arraybackslash}m{0.6cm}} \toprule
\multirow{4}{*}{State} & \multicolumn{4}{c}{Pedestrians} & \multicolumn{12}{c}{Vehicles} \\ \cmidrule{2-17}
& \multirow{3}{*}{$\text{P}_2$} & \multirow{3}{*}{$\text{P}_4$} & \multirow{3}{*}{$\text{P}_6$} & \multirow{3}{*}{$\text{P}_8$} & \multicolumn{3}{c}{$\text{V}_{\text{N}\rightarrow \text{S}}$} & \multicolumn{3}{c}{$\text{V}_{\text{S}\rightarrow \text{N}}$} & \multicolumn{3}{c}{$\text{V}_{\text{E}\rightarrow \text{O}}$} & \multicolumn{3}{c}{$\text{V}_{\text{O}\rightarrow \text{E}}$} \\ \cmidrule{6-17}
& &  & & & 5 & 2 & 12 & 1 & 6 & 16 & 7 & 4 & 14 & 3 & 8 & 18 \\
& & & & & \multicolumn{1}{c}{$\drsh$} & \multicolumn{1}{c}{$\downarrow$} & \multicolumn{1}{c}{$\dlsh$} & \multicolumn{1}{c}{$\Lsh$} & \multicolumn{1}{c}{$\uparrow$} & \multicolumn{1}{c}{$\Rsh$} & \multicolumn{1}{c}{\rotatebox[origin=c]{90}{$\Lsh$}}& \multicolumn{1}{c}{$\leftarrow$} & \multicolumn{1}{c}{\rotatebox[origin=c]{270}{$\dlsh$}} & \multicolumn{1}{c}{\rotatebox[origin=c]{90}{$\drsh$}} & \multicolumn{1}{c}{$\rightarrow$} & \multicolumn{1}{c}{\rotatebox[origin=c]{90}{$\dlsh$}} \\ \midrule
1 & \tikzcircle{3pt} & \tikzcircle{3pt} & \tikzcircle{3pt} & \tikzcircle{3pt} & \tikzcircle[green, fill=green]{3pt} & \tikzcircle{3pt} & \tikzcircle{3pt} & \tikzcircle[green, fill=green]{3pt} & \tikzcircle{3pt} & \tikzcircle{3pt} & \tikzcircle{3pt} & \tikzcircle{3pt} & \tikzcircle{3pt} & \tikzcircle{3pt} & \tikzcircle{3pt} & \tikzcircle{3pt} \\ \midrule
2 & \tikzcircle{3pt} & \tikzcircle{3pt} & \tikzcircle{3pt} & \tikzcircle{3pt} & \tikzcircle[yellow, fill=yellow]{3pt} & \tikzcircle{3pt} & \tikzcircle{3pt} & \tikzcircle[green, fill=green]{3pt} & \tikzcircle{3pt} & \tikzcircle{3pt} & \tikzcircle{3pt} & \tikzcircle{3pt} & \tikzcircle{3pt} & \tikzcircle{3pt} & \tikzcircle{3pt} & \tikzcircle{3pt} \\ \midrule
3 & \tikzcircle{3pt} & \tikzcircle{3pt} & \tikzcircle[green, fill=green]{3pt} & \tikzcircle{3pt} & \tikzcircle{3pt} & \tikzcircle{3pt} & \tikzcircle{3pt} & \tikzcircle[green, fill=green]{3pt} & \tikzcircle[green, fill=green]{3pt} & \tikzcircle[green, fill=green]{3pt} & \tikzcircle{3pt} & \tikzcircle{3pt} & \tikzcircle{3pt} & \tikzcircle{3pt} & \tikzcircle{3pt} & \tikzcircle{3pt} \\ \midrule
4 & \tikzcircle{3pt} & \tikzcircle{3pt} & \tikzcircle[green, fill=green]{3pt} & \tikzcircle{3pt} & \tikzcircle{3pt} & \tikzcircle{3pt} & \tikzcircle{3pt} & \tikzcircle[yellow, fill=yellow]{3pt} & \tikzcircle[green, fill=green]{3pt} & \tikzcircle[green, fill=green]{3pt} & \tikzcircle{3pt} & \tikzcircle{3pt} & \tikzcircle{3pt} & \tikzcircle{3pt} & \tikzcircle{3pt} & \tikzcircle{3pt} \\ \midrule
5 & \tikzcircle[green, fill=green]{3pt} & \tikzcircle{3pt} & \tikzcircle[green, fill=green]{3pt} & \tikzcircle{3pt} & \tikzcircle[green, fill=green]{3pt} & \tikzcircle[green, fill=green]{3pt} & \tikzcircle[green, fill=green]{3pt} & \tikzcircle{3pt} & \tikzcircle[green, fill=green]{3pt} & \tikzcircle[green, fill=green]{3pt} & \tikzcircle{3pt} & \tikzcircle{3pt} & \tikzcircle{3pt} & \tikzcircle{3pt} & \tikzcircle{3pt} & \tikzcircle{3pt} \\ \midrule
6 & \tikzcircle[green, fill=green]{3pt} & \tikzcircle{3pt} & \tikzcircle[green, fill=green]{3pt} & \tikzcircle{3pt} & \tikzcircle{3pt} & \tikzcircle[yellow, fill=yellow]{3pt} & \tikzcircle[yellow, fill=yellow]{3pt} & \tikzcircle{3pt} & \tikzcircle[yellow, fill=yellow]{3pt} & \tikzcircle[yellow, fill=yellow]{3pt} & \tikzcircle{3pt} & \tikzcircle{3pt} & \tikzcircle{3pt} & \tikzcircle{3pt} & \tikzcircle{3pt} & \tikzcircle{3pt} \\ \midrule
7 & \tikzcircle{3pt} & \tikzcircle{3pt} & \tikzcircle{3pt} & \tikzcircle{3pt} & \tikzcircle{3pt} & \tikzcircle{3pt} & \tikzcircle{3pt} & \tikzcircle{3pt} & \tikzcircle{3pt} & \tikzcircle{3pt} & \tikzcircle[green, fill=green]{3pt} & \tikzcircle{3pt} & \tikzcircle{3pt} & \tikzcircle[green, fill=green]{3pt} & \tikzcircle{3pt} & \tikzcircle{3pt} \\ \midrule
8 & \tikzcircle{3pt} & \tikzcircle{3pt} & \tikzcircle{3pt} & \tikzcircle{3pt} & \tikzcircle{3pt} & \tikzcircle{3pt} & \tikzcircle{3pt} & \tikzcircle{3pt} & \tikzcircle{3pt} & \tikzcircle{3pt} & \tikzcircle[yellow, fill=yellow]{3pt} & \tikzcircle{3pt} & \tikzcircle{3pt} & \tikzcircle[green, fill=green]{3pt} & \tikzcircle{3pt} & \tikzcircle{3pt} \\ \midrule
9 & \tikzcircle{3pt} & \tikzcircle{3pt} & \tikzcircle{3pt} & \tikzcircle{3pt} & \tikzcircle{3pt} & \tikzcircle{3pt} & \tikzcircle{3pt} & \tikzcircle{3pt} & \tikzcircle{3pt} & \tikzcircle{3pt} & \tikzcircle{3pt} & \tikzcircle{3pt} & \tikzcircle{3pt} & \tikzcircle[green, fill=green]{3pt} & \tikzcircle[green, fill=green]{3pt} & \tikzcircle[green, fill=green]{3pt} \\ \midrule
10 & \tikzcircle{3pt} & \tikzcircle{3pt} & \tikzcircle{3pt} & \tikzcircle{3pt} & \tikzcircle{3pt} & \tikzcircle{3pt} & \tikzcircle{3pt} & \tikzcircle{3pt} & \tikzcircle{3pt} & \tikzcircle{3pt} & \tikzcircle{3pt} & \tikzcircle{3pt} & \tikzcircle{3pt} & \tikzcircle[yellow, fill=yellow]{3pt} & \tikzcircle[green, fill=green]{3pt} & \tikzcircle[green, fill=green]{3pt} \\ \midrule
11 & \tikzcircle{3pt} & \tikzcircle{3pt} & \tikzcircle{3pt} & \tikzcircle{3pt} & \tikzcircle{3pt} & \tikzcircle{3pt} & \tikzcircle{3pt} & \tikzcircle{3pt} & \tikzcircle{3pt} & \tikzcircle{3pt} & \tikzcircle{3pt} & \tikzcircle[green, fill=green]{3pt} & \tikzcircle[green, fill=green]{3pt} & \tikzcircle{3pt} & \tikzcircle[green, fill=green]{3pt} & \tikzcircle[green, fill=green]{3pt} \\ \midrule
12 & \tikzcircle{3pt} & \tikzcircle[green, fill=green]{3pt} & \tikzcircle{3pt} & \tikzcircle[green, fill=green]{3pt} & \tikzcircle{3pt} & \tikzcircle{3pt} & \tikzcircle{3pt} & \tikzcircle{3pt} & \tikzcircle{3pt} & \tikzcircle{3pt} & \tikzcircle{3pt} & \tikzcircle[green, fill=green]{3pt} & \tikzcircle[green, fill=green]{3pt} & \tikzcircle{3pt} & \tikzcircle[green, fill=green]{3pt} & \tikzcircle[green, fill=green]{3pt} \\ \midrule
13 & \tikzcircle{3pt} & \tikzcircle[green, fill=green]{3pt} & \tikzcircle{3pt} & \tikzcircle[green, fill=green]{3pt} & \tikzcircle{3pt} & \tikzcircle{3pt} & \tikzcircle{3pt} & \tikzcircle{3pt} & \tikzcircle{3pt} & \tikzcircle{3pt} & \tikzcircle{3pt} & \tikzcircle[yellow, fill=yellow]{3pt} & \tikzcircle[yellow, fill=yellow]{3pt} & \tikzcircle{3pt} & \tikzcircle[yellow, fill=yellow]{3pt} & \tikzcircle[yellow, fill=yellow]{3pt} \\ \midrule 
\end{tabular}}
\end{table*}

\section{Implementation}\label{sec:implementation}
In order to evaluate the proposed system, we implemented the solution using BLE~\cite{bluetooth_v4_0} as the communication technology. This technology was chosen due to the solution requirements of low power, small size and low cost of modules and the compatibility with most mobile phones, tablets and other smart devices. Raspberry PIs\cite{pi2013raspberry} model B and the wiringPi library~\cite{henderson2013wiringpi} were used for the development of retransmitters and the controller. Regarding the Bluetooth modules, the chip module RN4020 from Microchip was used because of its small form factor, it has a Bluetooth low energy (version 4.1) stack on-board and it's controlled via simple ASCII AT commands over the UART interface~\cite{rn4020}. This module has a maximum announced range of $100~\text{m}$ using the chip's maximum power of $+7.5\text{dBm}$, making possible to use these modules to establish communications between personal smart devices, retransmitters and centralized controllers.

To implement a virtual traffic light system in a 4-way intersession, we started by enumerating all the possible phases, as depicted Table~\ref{tab:semaphore_states}, while permitting all combinations of movements and avoiding conflicts. Temporal states are represented by the numbers one to thirteen; traffic lights for every possible movement are represented by the colored circles. A red circle represents a prohibited movement, i.e. it must stop at the intersection, while green circles are used to permit the user movement. Finally, yellow circles indicate that a phase change is eminent. The numbering of user possible user movements, including pedestrians, follows the numbering model from~\cite{urbanik2015signal} depicted in Figure~\ref{fig:phases}.

\subsection{Controller}\label{sec:controller}
The controller entity is responsible for defining all the temporal states (Table ~\ref{tab:semaphore_states}). In this case the phases are statically defined and therefore, the controller continuously broadcasts the same information to all retransmitters and/or users. The controllers hardware is composed by a Module RN3020 connected to a Raspberry Pi Model B\cite{pi2013raspberry}.

Algorithm~\ref{alg:broadcast_messages} depicts the process of generating and broadcasting the traffic lights states. Firstly, the states' numbers are converted to American Standard Code for Information Interchange (ASCII) strings. A 16 bits Cyclic Redundancy Check (CRC)~\cite{4066263} is then performed in order to prevent an accidental propagation of errors and guarantee payload data integrity. After concluding this, an array of messages is built and transmitted; each message holds a traffic light state concatenated with its 16 bit CRC code. Finally, the controller keeps transmitting, in sequence, all traffic lights states with a predefined period. This period was only defined for testing purposes, it is also possible to configure independent periods for each traffic light state.

\begin{algorithm}[!htb]
\caption{Controller algorithm}
\label{alg:broadcast_messages}
\begin{algorithmic}
\FOR{$i=1$ to $\text{NUMBER\_OF\_STATES}$}
\STATE state = ITOA( i );
\STATE checksum = CRC\_16( state );
\STATE tx\_message[i] = CONCATENATE( state, checksum ); 
\ENDFOR
\WHILE{true}
\FOR{$i=1$ to $\text{NUMBER\_OF\_STATES}$}
\STATE current\_time = CLOCK\_GETTIME();
\STATE timeout = current\_time + controller\_period;
\WHILE{current\_time < timeout}
\STATE BROADCAST\_MESSAGE( tx\_message[i] );
\STATE current\_time = CLOCK\_GETTIME();
\ENDWHILE
\ENDFOR
\ENDWHILE
\end{algorithmic}
\end{algorithm}

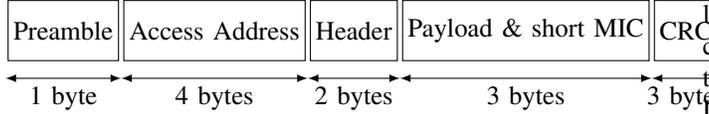
\begin{figure}[!htb]
\centering
\begin{tikzpicture}[scale=1, transform shape, node distance=0cm,outer sep=1pt,inner sep = 2pt]
	\tikzset{field/.style={align=left,shape=rectangle,minimum height=0.8cm,draw}}
	\tikzset{darkfield/.style={fill=gray!10,align=center,shape=rectangle,minimum height=0.8cm,draw}}

	\node [field]                      (preamble)     {Preamble};
	\node [field,right=of preamble] (access_addr)    {Access Address};
	\node [field,right=of access_addr]    (headers)    {Header};
	\node [field,right=of headers]    (payload)   {Payload \& short MIC};
	\node [field,right=of payload] (crc) {CRC};

	\draw [latex-latex] ($ (preamble.south west) - (0,0.2) $) -- node [auto,swap]
	{1 byte} ($ (preamble.south east) - (0,0.2) $);
	\draw [latex-latex] ($ (access_addr.south west) - (0,0.2) $) -- node [auto,swap]
	{4 bytes} ($ (access_addr.south east) - (0,0.2) $);
    \draw [latex-latex] ($ (headers.south west) - (0,0.2) $) -- node [auto,swap]
	{2 bytes} ($ (headers.south east) - (0,0.2) $);
    \draw [latex-latex] ($ (payload.south west) - (0,0.2) $) -- node [auto,swap]
	{3 bytes} ($ (payload.south east) - (0,0.2) $);
    \draw [latex-latex] ($ (crc.south west) - (0,0.2) $) -- node [auto,swap]
	{3 bytes} ($ (crc.south east) - (0,0.2) $);
\end{tikzpicture}
\caption{Traffic Lights BLE frame format}
\label{fig:frame}
\end{figure}

The format of the transmitted messages, including headers, is presented in Figure~\ref{fig:frame}. It was decided to apply a 16 bit CRC in detriment to the BLE's four bytes Message Integrity Check (MIC) in order to reduce the transmission time.

Transmitted message size is then thirteen bytes long. The time it takes to transmit one message is then given by the equation~\ref{eq:time_controller}.
\begin{align}
\label{eq:time_controller}
\frac{13~\text{bytes} * 8~\text{bits}}{1~\text{Mb}/\text{s}}=104\mu s
\end{align}
The Bluetooth Low Energy (BLE) standard~\cite{bluetooth_v4_0} specifies that when a device transmits one packet to a peer device, it should send back a packet with a minimum length, i.e. without payload and MIC, to confirm a successful packet reception. It also states that there is an mandatory Inter Frame Space ($\text{T\_ITS}$) equal to $150~\mu \text{s}$ . Thus we can evaluate controller data throughput as being:
\begin{align}
\text{Throughput}=\frac{(3*8)\text{bits}}{(80+150+104+150)\mu\text{s}}\approx0.050\text{b}/\mu\text{s}
\end{align}

\subsection{Retransmitter}\label{sec:retransmitter}

The main focus of this work was to provide a proof of concept, therefore a simple prototype of the retransmitter device was developed just for the relay of information. 
It locally stores the information provided by the controller and it has two different states: update and transmit. As the names imply, in the first state the device takes the role of observer listening to the data in the advertising packets sent by the controller. No connection happens between the controller and the retransmitter. Finally, it updates the stored information. During the second state, the retransmitter adds its ID to the message and broadcasts it to the medium.

The BLE network stack simplifies the process of configuring the network connection of a bluetooth device,  Generic Access Profile (GAP) provides a set of predefined methods to simplify the configuration. Using GAP a device can be configured as a broadcaster or as an observer. The algorithm~\ref{alg:retransmit_messages} depicts how the retransmitter uses GAP to perform the update and retransmissions of controller messages, it divides the controller's traffic lights broadcast period into two equal sized temporal slots.

On the first slot, it continuously receives messages and update the message to be retransmitted (tx\_msg) if the source Media Access Control (MAC) address equals the controller MAC address. On the second slot, it keeps retransmitting the current traffic light state if received by the controller otherwise, waits for the next controller period.

\begin{algorithm}[!htb]
\caption{Retransmitter algorithm}
\label{alg:retransmit_messages}
\begin{algorithmic}
\WHILE{(true)}
\STATE rx\_mac = 0;
\STATE message\_received = false;
\STATE current\_time = CLOCK\_GETTIME();
\STATE timeout = current\_time + (controller\_period / 2);
\STATE CONFIGURE\_BLE\_MODULE\_GAP( observer );
\WHILE{(current\_time < timeout)}
\STATE rx\_msg = RECEIVE\_MSG();
\STATE bool = VALIDATE\_CRC(rx\_msg);
\STATE rx\_mac = EXTRACT\_MAC( rx\_msg );
\IF{(bool == true) $\wedge$ (rx\_mac == mac\_controller)}
\STATE message\_received = true;
\STATE tx\_msg = rx\_msg;
\ENDIF
\STATE current\_time = CLOCK\_GETTIME();
\ENDWHILE
\IF{message\_received == true}
\STATE current\_time = CLOCK\_GETTIME();
\STATE timeout = current\_time + (controller\_period / 2);
\STATE tx\_msg = CONCATENATE( tx\_msg, retransmitter\_id );
\STATE CONFIGURE\_BLE\_MODULE\_GAP( broadcaster );
\WHILE{(current\_time < timeout)}
\STATE BROADCAST\_MESSAGE( tx\_msg );
\STATE current\_time = CLOCK\_GETTIME();
\ENDWHILE
\ELSE
\STATE WAIT( controller\_period / 2 );
\ENDIF
\ENDWHILE
\end{algorithmic}
\end{algorithm}

\subsection{User Interfaces}
The solution was tested using two different user interfaces. Instead of using one for the driver and the other for a pedestrian we used both to emulate the behaviour of the driver signalling device. This decision was made, to compare the performance of the smartphone with a dedicate BLE module. Thus, the first user interface was tested it a Samsung Galaxy TabA SM-T285 running the Android 5.1.1. The developped android application scans nearby BLE devices, interprets the received information and displays it to the user. For the experimental tests we reduced the latency for the scan, which also increased the duty-cycle and consequently the battery drainage. Five states were implemented orange when anomalies where detected, yellow for status changing, green and red, for protected and prohibited movements and green with a crosswalk when vehicles are permitted to move but have to yield to passengers. 

The Dedicated BLE module connected to a Raspberry Pi works in a similar fashion to the retransmitter. The scanning status is persistent and upon a phase modification it changes the stored state. This system could be implemented as an Human-Machine Interface (HMI) providing information to the dashboard. Both interfaces were deployed inside a vehicle and tested in the same conditions. 
javascript:void(0);\section{Experimental Evaluation}\label{sec:experimental_eval}

In order to validate the operation of the proposed system, two sets of experimental tests were conducted in the field. Firstly, the communications link performance between the centralized controller and the BLE retransmitter was evaluated. For each test run, both devices were statically positioned in specified road locations. The distance between them was varied from 0 to 140 meters, and a set of different parameters was measured for each distance value, e.g. packet success rate (PSR), received signal strength indication (RSSI) and the update time of the VTL system's status.

\begin{figure}[!htb]
	\centering
	\includegraphics[width=0.9\columnwidth]{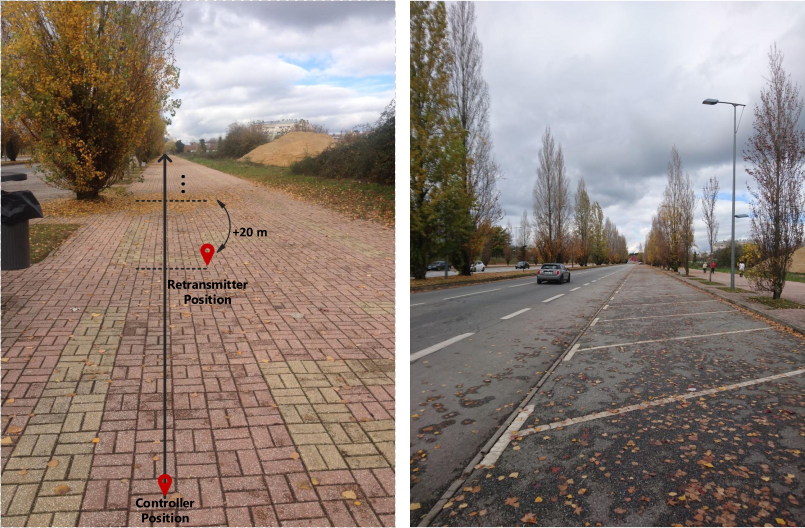}
	\caption{Evaluation of the BLE communications link performance between the centralized controller and the retransmitter device.}
	\label{fig:avenidaEuropa}
\end{figure}

\begin{figure}[!htb]
	\centering
	\includegraphics[width=0.9\columnwidth]{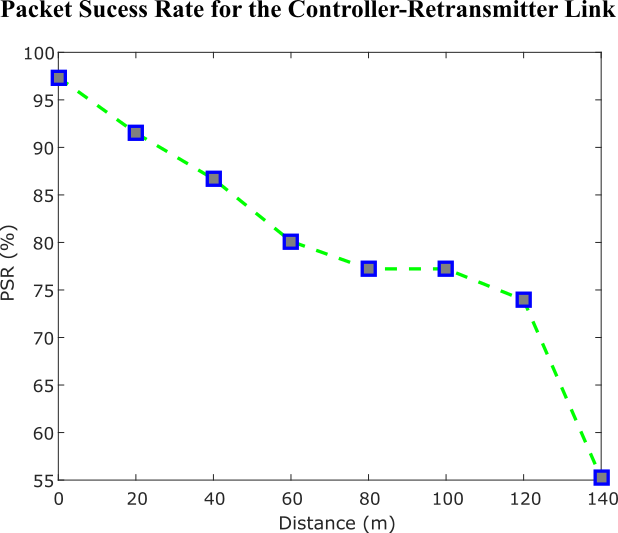}
	\caption{PSR as a function of the distance between the controller and the retransmitter.}
    \label{fig:psrStatic}
\end{figure}

This first set of static measurements were carried out at Avenida Europa, Viseu, Portugal, as depicted in figure \ref{fig:avenidaEuropa}. Packets were periodically transmitted during 20 seconds for each distance value and the number of correctly received packets was recorded in the retransmitter device. These results for the PSR parameter are presented in figure \ref{fig:psrStatic}.

\begin{figure}[!htb]
	\centering
	\includegraphics[width=\columnwidth]{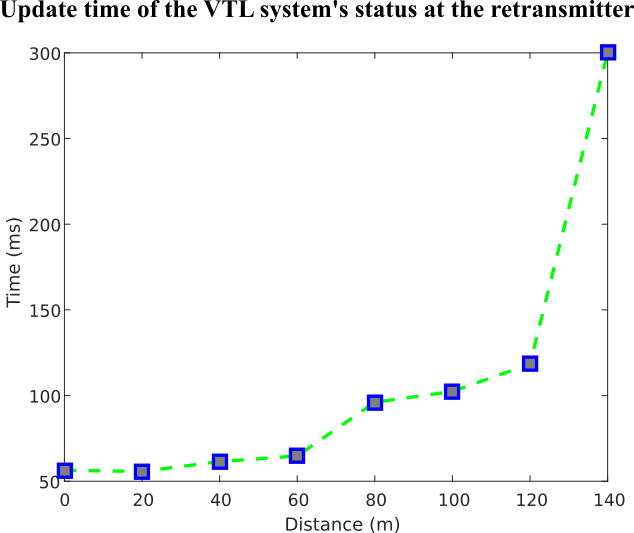}
	\caption{Update time of the VTL system's status at the retransmitter.}
	\label{fig:utStatic}
\end{figure}

Another important metric is the average update time of the VTL system's status at the retransmitter device. These measurements are displayed in figure \ref{fig:utStatic}. At short distances, the update time value is stable and approximately equal to 50 milliseconds. However, with the increasing distance, some packets start to be lost in the wireless medium and the system status can only be updated with the next message correctly received. As it can be observed in both figures \ref{fig:avenidaEuropa} and \ref{fig:utStatic}, the performance of the BLE link is reasonably good (PSR above 80\% and update time values below 100 ms) for distances lower than $\approx~60$ meters.

\begin{figure}[!htb]
	\centering
	\includegraphics[width=0.9\columnwidth]{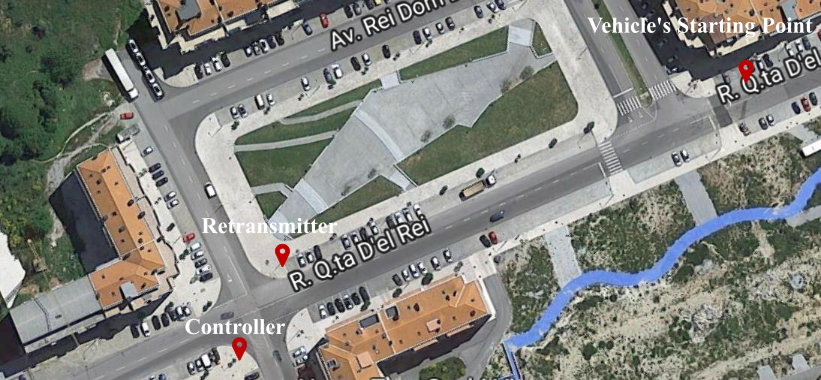}
	\caption{Experimental tests of the VTL system in an intersection with a moving vehicle.}
	\label{fig:intersect}
\end{figure}

The second set of experiments was performed with the goal of evaluating the system's operation in a close to the real-world scenario. In this case, a moving vehicle was approaching an intersection and receiving information from the VTL retransmitter. These experimental tests were conducted in an urban area with significant traffic flow, at Rua Quinta d'el Rei, in the city center of Viseu, Portugal (figure \ref{fig:intersect}). In this test, the distance between the centralized controller and the retransmitter was 37 meters, with no problems in the communications link, which presented good PSR and small update times.

Inside the vehicle, two types of communication devices were receiving the retransmitter messages: a smartphone running an Android application; and a dedicated BLE module connected to a Raspberry Pi. Both systems were operating simultaneously and collecting the results for analysis.

\begin{figure*}[!htb]
	\centering
	\includegraphics[width=0.9\textwidth]{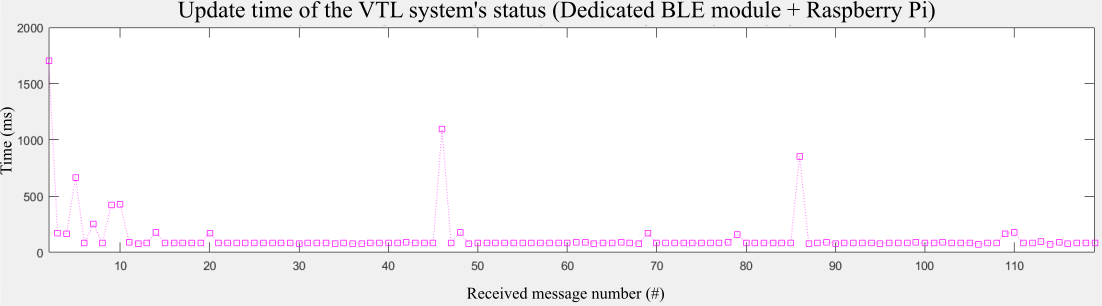}
	\caption{Update time of the VTL system's status in the Raspberry Pi.}
	\label{fig:dedicatedModule}
\end{figure*}

Figure \ref{fig:dedicatedModule} displays the obtained results for the update time in the Rapsberry Pi with respect to each successfully received message. The measurements are usually smaller than 100 ms, only with a few peaks when the number of messages lost between two consecutive well received packets increases.

\begin{figure*}[!htb]
	\centering
	\includegraphics[width=0.9\textwidth]{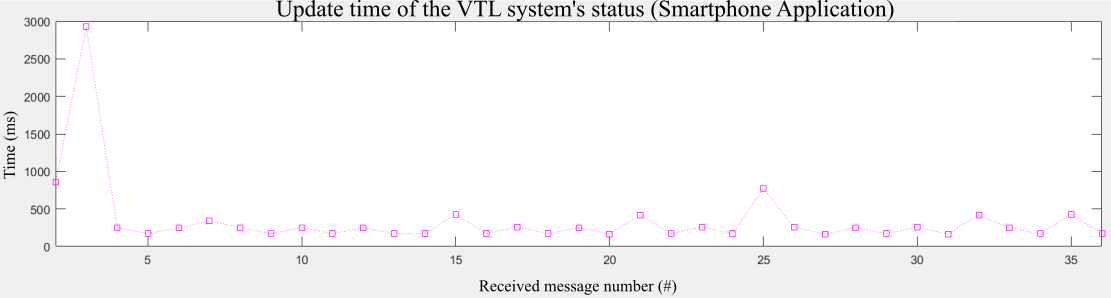}
	\caption{Update time of the VTL system's status in the Smartphone Application.}
	\label{fig:smartphoneDevice}
\end{figure*}

Figure \ref{fig:smartphoneDevice} presents the same results but for the case of the smartphone device. In this situation, the update times are significantly higher, with an average value clearly above 100 ms. It can also be observed that the number of successfully received messages in this case is much smaller ($\approx~35$ packets) than with the Raspberry Pi ($\approx~120$ packets). This result was expected due to the higher sensitivity of the dedicated BLE module.

It is possible to conclude from the obtained results that the proposed system can be implemented in a real-world scenario using BLE technology and personal devices (e.g. smartphone). However, in order to cope with the real-time demands of a safety-critical road traffic system, the performance of the personal VTL system can be greatly improved by employing dedicated hardware, such as the utilized BLE module plus the Raspberry Pi.

\section{Conclusions}
\label{sec:conclusions}
Going beyond the existing state-of-the-art, in this paper, we proposed a new architecture for virtual traffic light systems. The proposed architecture has unique features since it exploits the current ubiquity of smart devices, deploying a virtual traffic light system as close to the user as possible. Hence, users can be alerted to the current status of the nearest traffic sign, in the most convenient way, e.g, smartwatch display, phone vibration or even by a voice signal transmitted to their wireless headphones. The system was implemented using low cost, commercial-off-the-self hardware and tested in real-life scenarios.

As we explained before, in this paper we focused on providing a proof of concept for the proposed system. As future work, we aim to further explore the potentialities of the system. Specifically, we want to provide users with an active role, instead of passively receiving information. This step will unlock the adaptability of the overall system allowing the controllers to take better-informed decisions, improve traffic management and planning and ultimately provide a better user quality of experience. Finally, this solution is to be integrated in the PASMO open living lab for cooperative ITS and smart regions being deployed in \'Ilhavo, Aveiro (Portugal) \cite{ferreira2017pasmo}.

\section*{Acknowledgments}
This work is supported by the European Regional Development Fund (FEDER), through the Regional Operational Programme of Centre (CENTRO 2020) of the Portugal 2020 framework [Project PASMO with Nr. 000008 (CENTRO-01-0246-FEDER-000008)]. The work of João Almeida is supported by Fundação para a Ciência e Tecnologia - FCT/MEC through national funds under the PhD scholarship ref. SFRH/BD/52591/2014. 
% references section
\bibliography{bibliography.bib}
\bibliographystyle{IEEEtran}
\end{document}